\RequirePackage{fix-cm}
\documentclass[twocolumn,epjc3,final]{svjour3}  
\smartqed  
 \usepackage{mathrsfs}
\usepackage{amsmath, txfonts}
\usepackage{graphicx,bm,booktabs,bm}
\usepackage{longtable,lscape}
\usepackage{ulem} 
\usepackage{overpic,multirow,color}
\definecolor{cover}{rgb}{0.77,0.87,0.88}
\definecolor{blueone}{rgb}{0.1,0.1,.7}
\definecolor{citec}{rgb}{0.14,0.47,0.09}
\definecolor{two}{rgb}{0.0,0.5,0.}
\definecolor{three}{rgb}{.5,.1,0.15}
\usepackage[bookmarks=true,bookmarksopen=false,plainpages=false,breaklinks=true,
   bookmarksnumbered=true,hypertexnames=false,
   filecolor=blue,urlcolor=three,menucolor=three,
   linkcolor=three,citecolor=blueone, colorlinks,
   anchorcolor=blue,runcolor=pink,frenchlinks=red
   pdfstartview=FitH,pdftitle=title,%
   pdfauthor=author]{hyperref}
              \allowdisplaybreaks[4]

\graphicspath{{Figures/}} %

\allowdisplaybreaks[4]

\journalname{Eur. Phys. J. C}
\begin{document}
\title{Kaon-induced production of strange hidden-charm molecular pentaquarks $P^\Lambda_{\psi s}$ from proton}
\author{Shu Chen$^1$, Jun-Tao Zhu$^2$, Shu-Yi Kong$^2$ \and Jun He$^1$\thanksref{e1}
}                     
\thankstext{e1}{Corresponding author: junhe@njnu.edu.cn}
\institute{$^1$School of Physics and Technology, Nanjing
Normal University, Nanjing 210097, China\\
$^2$School of Microelectronics and Control Engineering, Changzhou University, Changzhou 213164, China}
\date{Received: date / Revised version: date}
%
\maketitle

\abstract{
In this work, we investigate the kaon-induced production of strange hidden-charm
pentaquark $P^\Lambda_{\psi s}$ states, including the $P^\Lambda_{\psi s}(4459)$ and $P^\Lambda_{\psi
s}(4338)$ observed by the LHCb collaboration. Coupled-channel interactions
involving the channels $\bar{K}N$, $\Xi_c^{(*)}\bar{D}^{(*)}$, $\Xi'_c\bar{D}^{(*)}$,
$\Lambda_c\bar{D}_s^{(*)}$, and $\Lambda J/\psi$ are studied using effective
Lagrangians to describe these interactions. The potential kernels are
constructed within the one-boson-exchange model and implemented into the
quasipotential Bethe-Salpeter equation to calculate the scattering amplitudes.
From these amplitudes, the partial-wave cross sections for kaon-induced
production are predicted alongside the poles corresponding
to strange hidden-charm molecular pentaquarks.  The analysis reveals complex
structures in the cross sections due to molecular states. A structure near the
$\Xi_c\bar{D}$ threshold is predicted in the $\Lambda_c\bar{D}_s$ channel,
corresponding to $P^\Lambda_{\psi s}(4338)$. Both molecular states near the
$\Xi_c\bar{D}^*$ threshold with $J^P = 1/2^-$ and $J^P = 3/2^-$, associated with
$P^\Lambda_{\psi s}(4459)$, produce peaks in the $\Xi'_c\bar{D}$ channel, which
vary significantly with the cutoff.  Additionally, a molecular state with $J^P =
3/2^-$ near the $\Xi_c^*\bar{D}$ threshold gives rise to a peak in the
$\Xi_c\bar{D}^*$ channel.  In the higher energy region, from 4350 to 4550 MeV,
although the cross sections in the $\Lambda_c\bar{D}_s^{(*)}$ channels are
substantial, no distinct structures corresponding to strange hidden-charm
pentaquarks are identified, apart from some threshold-related cusps.  These
results provide valuable theoretical insights and serve as a foundation for
future experimental studies of strange hidden-charm pentaquarks using kaon
beams at J-PARC and JLab.

 } 
\section{INTRODUCTION}

With the progress of experimental studies, particularly the observations at
LHCb, hidden-charm pentaquark states have garnered significant attention and
inspired extensive global research. The story began with the discovery of the
$P^N_{\psi}(4450)$ and $P^N_{\psi}(4380)$ states by the LHCb collaboration in
2015~\cite{Aaij:2015tga}, confirming predictions made by several
theorists~\cite{Wu:2010jy,Yang:2011wz,Wang:2011rga,Xiao:2013yca}. This
experimental breakthrough led to a surge of theoretical investigations into
hidden-charm pentaquark structures~\cite{Chen:2016qju,Guo:2017jvc,Chen:2015loa,Chen:2015moa,Karliner:2015ina,Roca:2015dva,He:2015cea,Burns:2015dwa}. The narrow widths and masses of these states, which are close to
the corresponding thresholds, strongly support the molecular state
interpretation. Specifically, $P^N_{\psi}(4380)$ and $P^N_{\psi}(4450)$ have
been assigned as $\bar{D}\Sigma^*_c$ and $\bar{D}^*\Sigma_c$ molecular states,
respectively, in our previous work~\cite{He:2015cea}.  In subsequent
observations, the $P^N_{\psi}(4450)$ state was resolved into two separate
structures, $P^N_{\psi}(4440)$ and $P^N_{\psi}(4457)$, and a new state,
$P^N_{\psi}(4312)$, was discovered in 2019~\cite{Aaij:2019vzc}. These states are
interpreted as S-wave molecular states composed of anticharmed mesons and
charmed baryons, with their masses closely aligned with the relevant
thresholds~\cite{Liu:2019tjn,Chen:2019asm,Xiao:2019aya,He:2019ify,Chen:2019bip,Du:2019pij}.
Given the success of the molecular picture in describing non-strange
hidden-charm systems, it is natural to extend this framework to hidden-charm
pentaquark states with strangeness~\cite{Anisovich:2015zqa,Wang:2015wsa,Feijoo:2015kts,Lu:2016roh,Chen:2015sxa,Chen:2016ryt,Xiao:2019gjd,Zhang:2020cdi,Wang:2019nvm}. The experimental observation of strange
hidden-charm pentaquarks is therefore pivotal in advancing our understanding of
molecular states with hidden charm.

In 2020, the LHCb Collaboration reported the observation of the strange
hidden-charm pentaquark $P^\Lambda_{\psi s}(4459)$ in the $\Xi_b^- \to J/\psi
\Lambda K^-$ decay~\cite{Aaij:2020gdg}. This state has a measured mass
approximately 19 MeV below the $\Xi_c \bar{D}^*$ threshold and a width of 17
MeV~\cite{Aaij:2020gdg}, characteristics consistent with its interpretation as a
molecular state. Numerous studies have investigated the molecular state nature
of $P^\Lambda_{\psi s}(4459)$.  For instance, our previous work suggests that
the structure of $P^\Lambda_{\psi s}(4459)$ primarily arises from the $\Xi_c
\bar{D}^*$ configuration with $J^P = 1/2^-$, while the possible contribution of
the $\Xi_c \bar{D}^*$ state with $J^P = 3/2^-$ cannot be
excluded~\cite{Zhu:2021lhd,Zhu:2022wpi}. Similarly, QCD sum rule analyses in
Ref.~\cite{Chen:2020uif} support the interpretation of $P^\Lambda_{\psi
s}(4459)$ as a $\Xi_c \bar{D}^*$ molecular state with $J^P = 1/2^-$ or $3/2^-$.
Ref.~\cite{Wang:2021itn} interprets $P^\Lambda_{\psi s}(4459)$ as either a
$\Xi'_c \bar{D}$ molecular state with quantum numbers $I(J^P) = 0(3/2^-)$ or a
$\Xi_c \bar{D}^*$ molecular state with $I(J^P) = 0(3/2^-)$. Moreover, studies
within the framework of heavy quark spin symmetry (HQSS), such as those in
Ref.~\cite{Peng:2020hql}, favor the $\Xi_c \bar{D}^*$ configuration with $J^P =
3/2^-$ as the most likely candidate for $P^\Lambda_{\psi s}(4459)$, ruling out
the $J^P = 1/2^-$ possibility. A recent study suggests that the strong coupling
between $\bar{D}_s \Lambda_c$ and $\bar{D} \Xi_c$, as well as between $\bar{D}s^*
\Lambda_c$ and $\bar{D}^*\Xi_c$, leads to the association of $P^\Lambda_{\psi
s}(4338)$ and $P^\Lambda_{\psi s}(4459)$ with states predominantly coupling to
$\bar{D} \Xi'$ and $\bar{D}^* \Xi_c$, respectively~\cite{Feijoo:2022rxf}.

In 2022, the LHCb Collaboration identified a new neutral strange hidden-charm
pentaquark state, $P^\Lambda_{\psi s}(4338)$, in the $B^- \to J/\psi \Lambda
\bar{p}$ decay~\cite{LHCb:2022ogu}. The state was measured to have a
mass of $4338.3 \pm 0.7 \pm 0.4$ MeV and a width of $7.0 \pm 1.2 \pm 1.3$ MeV.
The mass and narrow width of $P^\Lambda_{\psi s}(4338)$ are consistent with the
characteristics of a molecular state, which has inspired several
studies~\cite{Wang:2022neq,Wang:2022mxy,Yan:2022wuz,Ozdem:2022kei,Wang:2022tib,Nakamura:2022gtu,Giachino:2022pws,Ortega:2022uyu} proposing its
molecular nature.  In our previous work, the lineshape of the
$P^\Lambda_{\psi s}(4338)$ structure can be approximately reproduced by a narrow
molecular state resulting from the $\Xi_c \bar{D}$ interaction with $J^P =
1/2^-$, which lies very close to the threshold and exhibits a significant
interference effect~\cite{Zhu:2022wpi}.

Although hidden-charm pentaquarks, including those with strangeness, have been
observed at LHCb, further confirmation and a deeper understanding of these
states require analyzing their production mechanisms across different
 reaction channels.  Even before the observation of
the $P^N_\psi$ state at LHCb, the production of these states via
proton-antiproton collisions and photoproduction was proposed~\cite{Wu:2010jy,He:2015cea,Huang:2013mua}. Following the LHCb observations, many studies were
proposed to investigate the production of hidden-charm pentaquarks,
$P^\Lambda_{\psi}$, via pion-induced reactions or
photoproduction~\cite{Lu:2015fva,Huang:2016tcr,Kim:2016cxr,Cao:2019gqo,Wang:2019krd,Xie:2020niw,Shi:2022ipx,Wang:2019dsi}.  Additionally, the
pion-induced production of strange hidden-charm pentaquarks was studied even
before the LHCb observation~\cite{Wang:2015xwa}.  In Refs.~\cite{Clymton:2021thh,Paryev:2023icm,Paryev:2022zdx,Clymton:2022qlr}, the authors employed two
different methods, the effective Lagrangian approach and the Regge method, to
study the production of the exotic hidden-charm pentaquark state
$P^\Lambda_{\psi s}(4459)$ in the $\bar{K}p \rightarrow J/\psi \Lambda$
reaction. In Ref.~\cite{Cheng:2021gca}, the authors explored the possibility of
investigating the strange hidden-charm pentaquark state $P^\Lambda_{\psi
s}(4459)$ through photon-induced reactions on a proton target in the $\gamma p
\rightarrow K^+ P^\Lambda_{\psi s}(4459)$ channel, using an effective Lagrangian
approach.

In most studies of photoproduction or meson production of hidden-charm
pentaquarks, information about the pentaquarks is extracted directly from
experiments, and coupled-channel effects are not considered. In the current
work, we aim to provide a more comprehensive study of the production of
hidden-charm pentaquarks in molecular state picture through kaon-nucleon
scattering. Building on our previous research~\cite{Zhu:2021lhd,Zhu:2022wpi}, we calculate the coupled-channel
interactions between charmed baryons and anticharmed mesons within the framework
of the one-boson-exchange model.  In addition to the channels in the charmed
energy region, including $\bar{K}N$, $\Xi_c^{(*)}\bar{D}^{(*)}$,
$\Xi'_c\bar{D}^{(*)}$, $\Lambda_c\bar{D}_s^{(*)}$, and $\Lambda J/\psi$, which
successfully reproduce the experimentally observed $P^\Lambda_{\psi s}(4459)$
and $P^\Lambda_{\psi s}(4438)$~\cite{Zhu:2022wpi}, we extend our study by incorporating the light
channel $\bar{K}N$ for kaon-induced production. Using the quasipotential
Bethe-Salpeter equation (qBSE), we calculate the scattering amplitudes, which can then
be used to determine the cross sections for the transition from the $\bar{K}N$ channel
to those in the charmed energy region.

This article is organized as follows. After the introduction,
Sec.~\ref{Sec: Formalism} outlines the Lagrangians used to construct the
potentials for coupled-channel interactions, the qBSE approach, and the methodology for calculating the scattering
cross sections. In Sec.~\ref{sec3}, we present the results for the cross sections
of kaon-induced production. Finally, Sec.~\ref{sum} provides a summary of the
entire work.

\section{Formalism}\label{Sec: Formalism}

In the present work, we calculate the cross sections by using coupled-channel
interactions to obtain the scattering amplitudes for the $\bar{K}N$ and charmed
channels within the quasipotential Bethe-Salpeter equation (qBSE) approach. To
start, the potential kernel is derived from effective Lagrangians within the
one-boson-exchange model.

\subsection{Relevant lagrangians and potential kernels}

In the present work, we consider molecular pentaquarks produced from the
interaction of charmed baryons and anticharmed mesons, including the channels
$\Xi_c^*\bar{D}^*$, $\Xi'_c\bar{D}^*$, $\Xi^*_c\bar{D}$, $\Xi_c\bar{D}^*$,
$\Xi'_c\bar{D}$, $\Lambda_c\bar{D}_s$, $\Xi_c\bar{D}$, and $\Lambda_c\bar{D}_s$,
with light meson exchange within the one-boson-exchange model. We also consider
the channel involving a hidden-charm meson, $\Lambda J/\psi$, for which the
interaction is neglected. Therefore, we require vertices for the
coupling between the constituent particles and the exchanged meson. The
relevant Lagrangians for these vertices are given as follows~\cite{Zhu:2021lhd,Cheng:1992xi,Yan:1992gz,Wise:1992hn,Casalbuoni:1996pg},
\begin{align}
  \mathcal{L}_{\mathcal{\tilde{P}}^*\mathcal{\tilde{P}}\mathbb{P}} &=
 i\frac{2g\sqrt{m_{\mathcal{\tilde{P}}} m_{\mathcal{\tilde{P}}^*}}}{f_\pi}
  (-\mathcal{\tilde{P}}^{*\dag}_{a\lambda}\mathcal{\tilde{P}}_b
  +\mathcal{\tilde{P}}^\dag_{a}\mathcal{\tilde{P}}^*_{b\lambda})
  \partial^\lambda\mathbb{P}_{ab},\nonumber\\
    \mathcal{L}_{\mathcal{\tilde{P}}^*\mathcal{\tilde{P}}^*\mathbb{P}} &=
-\frac{g}{f_\pi} \epsilon_{\alpha\mu\nu\lambda}\mathcal{\tilde{P}}^{*\mu\dag}_a
\overleftrightarrow{\partial}^\alpha \mathcal{\tilde{P}}^{*\lambda}_{b}\partial^\nu\mathbb{P}_{ba},\nonumber\\
    \mathcal{L}_{\mathcal{\tilde{P}}^*\mathcal{\tilde{P}}\mathbb{V}} &=
\sqrt{2}\lambda g_V\varepsilon_{\lambda\alpha\beta\mu}
  (-\mathcal{\tilde{P}}^{*\mu\dag}_a\overleftrightarrow{\partial}^\lambda
  \mathcal{\tilde{P}}_b  +\mathcal{\tilde{P}}^\dag_a\overleftrightarrow{\partial}^\lambda
 \mathcal{\tilde{P}}_b^{*\mu})\partial^{\alpha}\mathbb{V}^\beta_{ab},\nonumber\\
	\mathcal{L}_{\mathcal{\tilde{P}}\mathcal{\tilde{P}}\mathbb{V}} &= i\frac{\beta	g_V}{\sqrt{2}}\mathcal{\tilde{P}}_a^\dag
	\overleftrightarrow{\partial}_\mu \mathcal{\tilde{P}}_b\mathbb{V}^\mu_{ab}, \nonumber\\
  \mathcal{L}_{\mathcal{\tilde{P}}^*\mathcal{\tilde{P}}^*\mathbb{V}} &= - i\frac{\beta
  g_V}{\sqrt{2}}\mathcal{\tilde{P}}_a^{*\dag}\overleftrightarrow{\partial}_\mu
  \mathcal{\tilde{P}}^*_b\mathbb{V}^\mu_{ab}\nonumber\\
  &-i2\sqrt{2}\lambda  g_Vm_{\mathcal{\tilde{P}}^*}\mathcal{\tilde{P}}^{*\mu\dag}_a\mathcal{\tilde{P}}^{*\nu}_b(\partial_\mu\mathbb{V}_\nu-\partial_\nu\mathbb{V}_\mu)_{ab}
,\nonumber\\
  \mathcal{L}_{\mathcal{\tilde{P}}\mathcal{\tilde{P}}\sigma} &=
  -2g_s m_{\mathcal{\tilde{P}}}\mathcal{\tilde{P}}_a^\dag \mathcal{\tilde{P}}_a\sigma, \nonumber\\
  \mathcal{L}_{\mathcal{\tilde{P}}^*\mathcal{\tilde{P}}^*\sigma} &=
  2g_s m_{\mathcal{\tilde{P}}^*}\mathcal{\tilde{P}}_a^{*\dag}
  \mathcal{\tilde{P}}^*_a\sigma,\label{LD}
\end{align}
where  the $\mathcal{\tilde{P}}=(\bar{D}^0, D^-, D^-_s)$, and  the $\mathbb P$ and $\mathbb V$ are the pseudoscalar and vector matrices as
\begin{align}
    {\mathbb P}&=\left(\begin{array}{ccc}
        \frac{\sqrt{3}\pi^0+\eta}{\sqrt{6}}&\pi^+&K^+\\
        \pi^-&\frac{-\sqrt{3}\pi^0+\eta}{\sqrt{6}}&K^0\\
        K^-&\bar{K}^0&-\frac{2\eta}{\sqrt{6}}
\end{array}\right),
\mathbb{V}&=\left(\begin{array}{ccc}
\frac{\rho^{0}+\omega}{\sqrt{2}}&\rho^{+}&K^{*+}\\
\rho^{-}&\frac{-\rho^{0}+\omega}{\sqrt{2}}&K^{*0}\\
K^{*-}&\bar{K}^{*0}&\phi
\end{array}\right),\label{MPV}
\end{align}
where the indices $a, b=1, 2, 3$ are used to label the particle elements in the matrices ${\mathbb P}, {\mathbb V}$ and vector $\mathcal{P}$.

For the baryon side, the Lagrangians for the couplings between the charmed
baryons and light mesons are given as follows,
\begin{align}
{\cal L}_{BB\mathbb{P}}&=-i\frac{3g_1}{4f_\pi\sqrt{m_{\bar{B}}m_{B}}}~\epsilon^{\mu\nu\lambda\kappa}\partial_\nu \mathbb{P}~
\sum_{i,j=0,1}\bar{B}_{i\mu} \overleftrightarrow{\partial}_\kappa B_{j\lambda},\nonumber\\
{\cal L}_{BB\mathbb{V}}&=-\frac{\beta_S g_V}{2\sqrt{2m_{\bar{B}}m_{B}}}\mathbb{V}^\nu
 \sum_{i,j=0,1}\bar{B}_i^\mu \overleftrightarrow{\partial}_\nu B_{j\mu}\nonumber\\
&-\frac{\lambda_S
g_V}{\sqrt{2}}(\partial_\mu \mathbb{V}_\nu-\partial_\nu \mathbb{V}_\mu) \sum_{i,j=0,1}\bar{B}_i^\mu B_j^\nu,\nonumber\\
{\cal L}_{BB\sigma}&=\ell_S\sigma\sum_{i,j=0,1}\bar{B}_i^\mu B_{j\mu},\nonumber\\
    {\cal L}_{B_{\bar{3}}B_{\bar{3}}\mathbb{V}}&=-\frac{g_V\beta_B}{2\sqrt{2m_{\bar{B}_{\bar{3}}}m_{B_{\bar{3}}}} }\mathbb{V}^\mu\bar{B}_{\bar{3}}\overleftrightarrow{\partial}_\mu B_{\bar{3}},\nonumber\\
{\cal L}_{B_{\bar{3}}B_{\bar{3}}\sigma}&=i\ell_B \sigma \bar{B}_{\bar{3}}B_{\bar{3}},\nonumber\\
{\cal L}_{BB_{\bar{3}}\mathbb{P}}
    &=-i\frac{g_4}{f_\pi} \sum_i\bar{B}_i^\mu \partial_\mu \mathbb{P} B_{\bar{3}}+{\rm H.c.},\nonumber\\
{\cal L}_{BB_{\bar{3}}\mathbb{V}}    &=\frac{g_\mathbb{V}\lambda_I}{\sqrt{2m_{\bar{B}}m_{B_{\bar{3}}}}} \epsilon^{\mu\nu\lambda\kappa} \partial_\lambda \mathbb{V}_\kappa\sum_i\bar{B}_{i\nu} \overleftrightarrow{\partial}_\mu
   B_{\bar{3}}+{\rm H.c.},
   \label{LB}
\end{align}
where the Dirac spinor operators with label $i, j=0,1$ are defined as,
\begin{align}
{B}_{0\mu}&\equiv -\sqrt{\frac{1}{3}}(\gamma_{\mu}+v_{\mu})\gamma^{5}B; \, \,  \,  B_{1\mu}\equiv B^{*}_{\mu},\nonumber\\
{\bar{B}}_{0\mu}&\equiv\sqrt{\frac{1}{3}}\bar{B}\gamma^{5}(\gamma_{\mu}+v_{\mu});\, \,  \, \bar{B}_{1\mu}\equiv \bar{B}^{*}_{\mu},
\end{align}
and  the charmed baryon matrices are defined as
\begin{align}
B_{\bar{3}}&=\left(\begin{array}{ccc}
0&\Lambda^+_c&\Xi_c^+\\
-\Lambda_c^+&0&\Xi_c^0\\
-\Xi^+_c&-\Xi_c^0&0
\end{array}\right),\quad
B=\left(\begin{array}{ccc}
\Sigma_c^{++}&\frac{1}{\sqrt{2}}\Sigma^+_c&\frac{1}{\sqrt{2}}\Xi'^+_c\\
\frac{1}{\sqrt{2}}\Sigma^+_c&\Sigma_c^0&\frac{1}{\sqrt{2}}\Xi'^0_c\\
\frac{1}{\sqrt{2}}\Xi'^+_c&\frac{1}{\sqrt{2}}\Xi'^0_c&\Omega^0_c
\end{array}\right), \nonumber\\
B^*&=\left(\begin{array}{ccc}
\Sigma_c^{*++}&\frac{1}{\sqrt{2}}\Sigma^{*+}_c&\frac{1}{\sqrt{2}}\Xi^{*+}_c\\
\frac{1}{\sqrt{2}}\Sigma^{*+}_c&\Sigma_c^{*0}&\frac{1}{\sqrt{2}}\Xi^{*0}_c\\
\frac{1}{\sqrt{2}}\Xi^{*+}_c&\frac{1}{\sqrt{2}}\Xi^{*0}_c&\Omega^{*0}_c
\end{array}\right).\label{MBB}
\end{align}

The masses of the particles involved in the calculation are chosen based on the
recommended central values from the Review of Particle Physics
(PDG)~\cite{Tanabashi:2018oca}. The mass of the broad $\sigma$ meson is taken as
500 MeV. The coupling constants used in the calculations are listed in
Table~\ref{coupling}.

\renewcommand\tabcolsep{0.16cm}
\renewcommand{\arraystretch}{1.2}
\begin{table}[h!]
\caption{The coupling constants adopted in the
calculation, which are cited from the literature~\cite{Chen:2019asm,Liu:2011xc,Isola:2003fh,Falk:1992cx}. The $\lambda$ and $\lambda_{S,I}$ are in the units of GeV$^{-1}$. Others are in the units of $1$.
\label{coupling}}
\begin{tabular}{cccccccccccccccccc}\toprule[1pt]
$\beta$&$g$&$g_V$&$\lambda$ &$g_{s}$&$\ell_S$\\
$0.9$&$0.59$&$5.9$&$0.56$ &$0.76$&$6.2$\\\hline
$\beta_S$&$g_1$&$\lambda_S$ &$\beta_B$&$\ell_B$ &$g_4$&$\lambda_I$\\
$-1.74$&$-0.94$&$-3.31$&$-\beta_S/2$&$-\ell_S/2$&$3g_1/{(2\sqrt{2})}$&$-\lambda_S/\sqrt{8}$ \\
\bottomrule[1pt]
\end{tabular}
\end{table}

Based on the aforementioned Lagrangians, we can derive the interaction
mechanisms responsible for generating molecular strange hidden-charm
pentaquarks. To investigate their kaon production dynamics, it is crucial to
incorporate couplings between these exotic states and the $\bar{K}N$ channel. While
the broad energy region covered here suggests potential contributions from multiple
inelastic channels, our current focus remains on providing experimentally
testable predictions. Notably, these inelastic channels primarily influence
cross-section magnitudes rather than the fundamental production mechanisms.
Given the current absence of experimental data for these exotic states, our
theoretical predictions target order-of-magnitude estimates - a resolution level
sufficient for guiding initial experimental searches. Moreover, the essential
characteristics of cross-section behavior (excluding absolute normalization)
remain relatively insensitive to these coupling details due to the large mass gap. We therefore employ a
simplified one-boson exchange framework to model the crucial coupling between
the $\bar{K}N$ channel and charmed-sector final states, maintaining theoretical
consistency while preserving predictive capability.

 To write the potential for the $\bar{K}N$ interaction, the Lagrangians constructed using heavy quark and chiral symmetries are introduced and presented explicitly as follows~\cite{Zhu:2022fyb,Ronchen:2012eg},
\begin{align}
\mathcal{L}_{\bar{K}\bar{K}{\rho} }&=
ig_{\bar{K}\bar{K}\rho}\bar{K}^{\dag}
{\bm \tau}\cdot{\bm \rho}^{\mu}\overleftrightarrow{\partial}_{\mu}\bar{K},   \nonumber\\
\mathcal{L}_{\bar{K}\bar{K}{\omega} }&=
ig_{\bar{K}\bar{K}\omega}\bar{K}^{\dag}
{\omega}^{\mu}\overleftrightarrow{\partial}_{\mu}\bar{K},   \nonumber\\
\mathcal{L}_{\bar{K}\bar{K}\sigma }&=
-g_{\bar{K}\bar{K}\sigma}\bar{K}^{\dag} \sigma\bar{K}, \nonumber\\
 \mathcal{L}_{NN\pi}
&=-\frac{g_{NN\pi}}{m_{\pi}}\bar{N}\gamma^5\gamma^{\mu}{\bm \tau}
\cdot\partial_{\mu}{\bm \pi}N,\quad\nonumber\\
\mathcal{L}_{NN\rho} &=
-g_{NN\rho}\bar{N}[\gamma^{\mu}-\frac{\kappa_{\rho}}{2m_{N}}\sigma^{\mu\nu}\partial_{\nu}]
{\bm \tau}\cdot{\bm \rho_{\mu}}N,
\nonumber\\
\mathcal{L}_{NN\omega} &=
-g_{NN\omega}\bar{N}[\gamma^{\mu}-\frac{\kappa_{\omega}}{2m_{N}}\sigma^{\mu\nu}\partial_{\nu}]
{ \omega_{\mu}}N,
\nonumber\\
\mathcal{L}_{NN\sigma}& =-g_{NN\sigma}\bar{N}N\sigma,
\end{align}
where $\bar{K}$, $N$, $\pi$, and $\rho$ represent the $\bar{K}$ meson, nucleon, pion meson, and $\rho$ meson fields, respectively. The coupling constants are chosen as follows: $g_{\bar{K}\bar{K}\sigma} = 3.65$, $g_{NN\pi} = 0.989$, $g_{NN\rho} = -3.1$, $\kappa_{\rho} = 1.825$, $\kappa_{\omega} = 0$, and $g_{NN\sigma} = 5$~\cite{Matsuyama:2006rp,Lu:2020qme}.

When calculating the couplings between $\bar{K}N$ and other channels,
in addition to the
Lagrangians mentioned above, we also need to use the following
Lagrangians~\cite{Wang:2016fhj,Okubo:1975sc,Sibirtsev:2000aw,Liu:2001ce,Dong:2009tg,Lu:2016nnt},
\begin{align}
{\cal L}_{DN\Lambda_c}&=ig_{DN\Lambda_c}(\bar{N}\gamma_5\Lambda_cD+\bar{D}\bar{\Lambda}_c\gamma_5N),\nonumber\\
{\cal L}_{D^*N\Lambda_c}&=g_{D^*N\Lambda_c}(\bar{N}\gamma_\mu\Lambda_cD^{*\mu}+\bar{D}^{*\mu}\bar{\Lambda}_c\gamma_{\mu}N),\nonumber\\
{\cal L}_{\Sigma_c N D^*} &=g_{\Sigma_c N D^*} \bar{N} \gamma_{\mu} {\vec \tau} \cdot {\vec \Sigma_c} D^{*\mu}+ \rm{H.c}.,\nonumber\\
{\cal L}_{\Sigma_c N D} &= -i g_{\Sigma_c N D} \bar{N} \gamma_5 {\vec \tau} \cdot {\vec \Sigma_c} D+ \rm{H.c}.\label{LB}
\end{align}
The coupling constants, $g_{DN\Lambda_c}$ and $g_{D^N\Lambda_c}$, are chosen as
10.7 and -5.8, respectively~\cite{Khodjamirian:2011sp}. Additionally,
$g_{\Sigma_c N D^*} = 3.0$~\cite{Dong:2009tg} and $g_{\Sigma_c N D} =
2.69$~\cite{Garzon:2015zva}.

In the current framework, the inclusion of nine  reaction channels naturally leads to 81  pairwise interactions requiring calculation. Explicitly enumerating all these interaction terms would not only prove exceptionally tedious but also introduce significant error-prone risks in formulation.
Following the method in Ref.~\cite{He:2019rva}, we input the vertices $\Gamma$ and propagators $P$ directly into the code. The explicit forms of the potential can be written using the Lagrangians and flavor wave functions as follows,
\begin{equation}%
{\cal V}_{\mathbb{P},\sigma}=f_I\Gamma_1\Gamma_2 P_{\mathbb{P},\sigma}f(q^2),\ \
{\cal V}_{\mathbb{V}}=f_I\Gamma_{1\mu}\Gamma_{2\nu}  P^{\mu\nu}_{\mathbb{V}}f(q^2).\label{V}
\end{equation}
The propagators are defined as usual as
\begin{equation}%
P_{\mathbb{P},\sigma}= \frac{i}{q^2-m_{\mathbb{P},\sigma}^2},\ \
P^{\mu\nu}_\mathbb{V}=i\frac{-g^{\mu\nu}+q^\mu q^\nu/m^2_{\mathbb{V}}}{q^2-m_\mathbb{V}^2}.
\end{equation}

The form factor $f(q^2)$ is introduced to account for the off-shell effects of
the exchanged meson. It is expressed as $f(q^2) = e^{-(m_e^2 - q^2)^2 /
\Lambda_e^2}$, where $m_e$ and $q$ denote the mass and momentum of the exchanged
meson, respectively, while $\Lambda_e$ serves as a cutoff parameter to mitigate
the on-shell effects. In the meson propagator, we substitute $q^2 \to -|q|^2$ to
eliminate singularities, following the method outlined in
Ref.~\cite{Gross:2008ps}. The factor $f_I$ represents the flavor contribution for
a given meson exchange in an isoscalar interaction, with the specific values
provided in Table~\ref{flavor factor}.

\renewcommand\tabcolsep{0.3cm}
\renewcommand{\arraystretch}{1.5}
\begin{table}[h!]
\caption{The flavor factors $f_I$ for certain meson exchanges of certain isoscalar interaction.
 \label{flavor factor}}
 \begin{tabular}{c|cccccc}\bottomrule[1pt]
  $$ &$\pi$&$\eta$& $\rho$ &$\omega$&$\sigma$  \\\hline
 $ \bar{K}N\rightarrow \bar{K}N$&$ $&$ $&$-\frac{3\sqrt{2}}{2}$&$\frac{\sqrt{2}}{2}$&$1$\\
 $\Xi_c \bar{D}^{(*)}\rightarrow \Xi_c \bar{D}^{(*)}$&$ $&$ $&$-\frac{3}{2}$&$\frac{1}{2} $ & $2$\\
 $\Xi_c^{(',*)} \bar{D}\rightarrow \Xi_c^{(',*)} \bar{D}$&$ $&$ $&$-\frac{3}{4}$&$\frac{1}{4}$ &$1$\\
 $\Xi_c^{(',*)} \bar{D}^{*}\rightarrow \Xi_c^{(',*)} \bar{D}^{*}$&$-\frac{3}{4}$&$-\frac{1}{12} $&$-\frac{3}{4}$&$\frac{1}{4}$ &$1$\\
 \bottomrule[1pt]
   $$ &$\Lambda_c $&$\Sigma_c^{(*)}$&$\bar{D}^{(*)}$  \\\hline
 $\bar{K}N\rightarrow \Xi_c^{(',*)}\bar{D^{(*)}}$&$-\frac{\sqrt{2}}{2}$&$-\frac{1}{2}$  \\
 $\bar{K}N\rightarrow \Xi_c\bar{D}^{(*)}$&$ $&$\frac{\sqrt{2}}{2}$   \\
 $ \bar{K}N\rightarrow \Lambda_c \bar{D_s}^{(*)}$&$ $&$ $&$\sqrt{2}$ \\
 \bottomrule[1pt]
 &$D$ & $ D^* $ & $D_s $& $D_s^* $&$  $ \\\hline
 $\bar{D}^{(*)}\Xi_c\to J/\psi\Lambda$  &$ -\sqrt{2} $&$-\sqrt{2}  $& &$$ & $$\\
 $\bar{D}^{(*)}\Xi^{(',*)}_c\to J/\psi\Lambda$  &$ -\sqrt{2}$&$-\sqrt{2} $&$$ &$$ & $$\\
 $\bar{D}_s^{(*)}\Lambda_c\to J/\psi\Lambda$  &$$ &$$ &$1 $&$1 $ & $$\\

 \toprule[1pt]
 \end{tabular}
\end{table}

\subsection{The qBSE approach}

By inserting the potential kernel into the Bethe-Salpeter equation, the
scattering amplitude is derived. The original 4-dimensional integral equation in
Minkowski space is simplified to a 1-dimensional integral equation through
partial-wave decomposition and the spectator quasipotential approximation, with
a fixed spin-parity $J^P$, allowing the scattering
amplitude to be calculated as~\cite{He:2015mja,He:2014nya,He:2012zd},
\begin{align}
i{\cal M}^{J^P}_{\lambda'\lambda}({\rm p}',{\rm p})
&=i{\cal V}^{J^P}_{\lambda',\lambda}({\rm p}',{\rm
p})+\sum_{\lambda''}\int\frac{{\rm
p}''^2d{\rm p}''}{(2\pi)^3}\nonumber\\
&\cdot
i{\cal V}^{J^P}_{\lambda'\lambda''}({\rm p}',{\rm p}'')
G_0({\rm p}'')i{\cal M}^{J^P}_{\lambda''\lambda}({\rm p}'',{\rm
p}),\quad\quad \label{Eq: BS_PWA}
\end{align}
where the summation is restricted to nonnegative helicities $\lambda''$. 
The propagator $G_0({\rm p}'')$ is simplified from its original four-dimensional form under the quasipotential approximation and takes the form
\begin{align} G_0 &= \frac{\delta^+(p''^{~2}_h - m_h^{2})}{p''^{2}_l - m_l^{2}} \nonumber\\ &= \frac{\delta^+(p''^{0}_h - E_h({\rm p}''))}{2E_h({\rm p}'')[(W - E_h({\rm p}''))^2 - E_l^2({\rm p}'')]}. \end{align} As required by the spectator approximation adopted in this work, the heavier particle (denoted as $h$) in a given channel is placed on shell~\cite{Gross:1999pd}, which satisfies $p''^0_h = E_h({\rm p}'') = \sqrt{m_h^2 + {\rm p}''^2}$. The energy of the lighter particle (denoted as $l$) is then given by $p''^0_l = W - E_h({\rm p}'')$.
Here and hereafter, the value of the momentum  in center-of-mass frame is defined as ${\rm p}=|{\bm p}|$.
  The partial wave potential is expressed as:
\begin{align}
{\cal V}_{\lambda'\lambda}^{J^P}({\rm p}',{\rm p})
&=2\pi\int d\cos\theta
~[d^{J}_{\lambda\lambda'}(\theta)
{\cal V}_{\lambda'\lambda}({\bm p}',{\bm p})\nonumber\\
&+\eta d^{J}_{-\lambda\lambda'}(\theta)
{\cal V}_{\lambda'-\lambda}({\bm p}',{\bm p})],
\end{align}
where $\eta = PP_1P_2(-1)^{J-J_1-J_2}$, with $P$ and $J$ denoting the parity and
spin of the system, as well as those of constituent particles 1 and 2. The
relative momenta for the initial and final states are specified as ${\bm
p}=(0,0,{\rm p})$ and ${\bm p}'=({\rm p}'\sin\theta,0,{\rm p}'\cos\theta)$,
respectively. The term $d^J_{\lambda\lambda'}(\theta)$ represents the Wigner
d-matrix.

To solve the integral equation (\ref{Eq: BS_PWA}), the momenta ${\rm p}$, ${\rm p}'$, and ${\rm p}''$ are discretized using the Gauss quadrature method with weights $w({\rm p}_i)$. The discretized form of the qBSE can then be expressed as~\cite{He:2015mja},
\begin{align}
{M}_{ik}
&={V}_{ik}+\sum_{j=0}^N{ V}_{ij}G_j{M}_{jk}.\label{Eq: matrix}
\end{align}
The propagator $G$ is of a form
\begin{align}
	G_{j>0}&=\frac{w({\rm p}''_j){\rm p}''^2_j}{(2\pi)^3}G_0({\rm
	p}''_j), \nonumber\\
G_{j=0}&=-\frac{i{\rm p}''_{\rm o}}{32\pi^2 W}+\sum_j
\left[\frac{w({\rm p}_j)}{(2\pi)^3}\frac{ {\rm p}''^2_{\rm o}}
{2W{({\rm p}''^2_j-{\rm p}''^2_{\rm o})}}\right].
\end{align}
The  ${\rm p}''_{\rm o}=\lambda^{1/2}(W,M_1,M_2)/2W$ is on-shell momentum with $\lambda(x,y,z)=[x^2-(y+z)^2][x^2-(y-z)^2]$ and $W$ represents the total
energy of the two-constituent system. To regularize the propagator, an
exponential form factor is introduced as $G_0({\rm p}'') \to G_0({\rm
p}'')\left[e^{-(p''^2_l-m_l^2)^2/\Lambda_r^4}\right]^2$, where $\Lambda_r$
denotes the cutoff~\cite{He:2015mja}. In our framework, the cutoffs applied in
the form factors, including those in the propagator and exchange interactions,
are treated as free parameters, with $\Lambda_e = \Lambda_r = \Lambda$. We adopt
a cutoff value of $\Lambda = 1.04$ GeV, which was successfully used to fit the
LHCb data in our previous work~\cite{Zhu:2022wpi}, allowing for slight variations of 0.1 GeV
to estimate uncertainties.

The poles of the nine-channel scattering amplitudes for different spin-parity
combinations $J^P$ are located by varying $z$ in the complex energy plane to
satisfy the condition $|1-V(z)G(z)|=0$. For a calculation involving $N$
channels, there are $2^N$ Riemann sheets associated with unitarity. In this
study, we adopt the method outlined in Ref.~\cite{Roca:2005nm} to identify the
poles. Specifically, the default propagator $G(z)$ is used for the energy region below the threshold, while $G(z) + i{\rm p}''_o / (16\pi^2 z)$ is adopted for the region above the threshold, corresponding to the first and second Riemann sheets, respectively. This treatment does not affect the physical results on the real energy axis and yields pole positions and half-widths that are closer to those of the corresponding Breit-Wigner forms.

The total cross section is calculated as~\cite{He:2015cca},
\begin{align}
    \sigma=\frac{1}{16\pi s}\frac{{\rm p}^\prime}{\rm p}\sum_{J^P,\lambda'\geq0\lambda\geq0}\frac{\tilde{J}}{\tilde{j_1}\tilde{j_2}}\left|\frac{M^{J^P}_{\lambda'\lambda}({\rm p}', {\rm p})}{4\pi}\right|^2.
\end{align}
Here, $s$ represents the square of the invariant mass of the initial particle system. The variables $J$ and $j_i$ denote the total angular momentum of the system and the spins of the two initial particles, respectively.

\section{Numerical Result}\label{sec3}

\subsection{Molecular states without coupled-channel effects}

In this section, we study the kaon-induced production of strange hidden-charm
pentaquarks, through considering the coupled-channel interactions. However, before
conducting a coupled-channel calculation, we first present the results without
coupled-channel effects for reference.
We analyze the interactions $\bar{K}N$, $\Xi_c^{(*)}\bar{D}^{(*)}$, $\Xi'_c\bar{D}^{(*)}$,
$\Lambda_c\bar{D}_s^{(*)}$, and $\Lambda J/\psi$, while turning off
the couplings between these channels. Under this setup, the poles in the complex
energy plane correspond directly to the single-channel interaction of each
respective channel.

\begin{figure}[h!]
  \centering
  \includegraphics[scale=0.72,bb=80 40 440 320,clip]{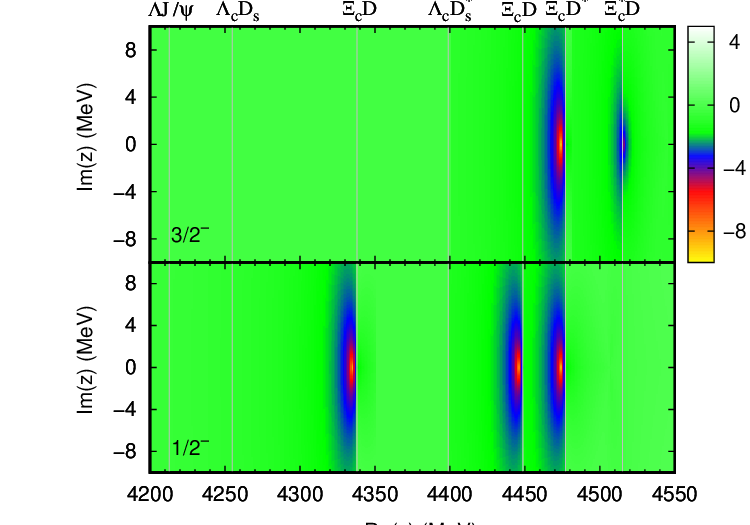}\\
  \caption{Poles for molecular $P^{\Lambda}_{\psi s}$ state with spin parities $3/2^-$ (upper panel) and $1/2^-$ (lower
  panel) obtained without considering coupled-channel effects. The colorboxe represent the
  values of $\log|1-V(z)G(z)|$. The gray lines correspond to the thresholds
  of $\Xi^*_c\bar{D}$, $\Xi_c\bar{D}^*$, $\Xi'_c\bar{D}$,
  $\Lambda_c\bar{D}_s^{*}$, $\Xi_c\bar{D}$,  $\Lambda_c\bar{D}_s$,
  and $\Lambda J/\psi$, from right to left.}\label{1}
\end{figure}

The parameters of the Lagrangians employed here are the same as those used in
our previous study on the strange hidden-charm pentaquark states based on LHCb
data, specifically $P^\Lambda_{\psi s}(4459)$ and $P^\Lambda_{\psi
s}(4338)$~\cite{Zhu:2022wpi}. The results are shown in Fig.~\ref{1}, where the
horizontal axis represents the real part of the pole position, and the vertical
axis denotes the imaginary part. In this work, only states with spin parity producible from
S-wave interactions are considered. 

Five poles are identified within the energy range of 4200 to 4550 MeV based on
the interactions considered. All poles identified in this work are located below threshold, meaning they reside on the Riemann sheet corresponding to the default propagator $G(z)$ of the respective channel. Since the $\bar{K}N$ threshold is significantly
below the charmed energy region and the couplings between channels are turned
off, the results closely resemble those from our previous work where the
$\bar{K}N$ channel was excluded~\cite{Zhu:2022wpi}. These poles correspond to five molecular states
generated from the single-channel interactions of $\Xi_c^*\bar{D}$ with spin
parity $J^P = 3/2^-$, $\Xi_c\bar{D}^*$ with $1/2^-$ and $3/2^-$, $\Xi'_c\bar{D}$
with $1/2^-$, and $\Xi_c\bar{D}$ with $1/2^-$. As the cutoff increases, all
poles shift further away from their respective thresholds. Two states with spin
parities $1/2^-$ and $3/2^-$ from the $\Xi_c\bar{D}^*$ interaction have nearly
the same mass as those found in the calculation without coupled-channel effects,
approximately 4474 MeV, which is slightly higher than the mass of
$P^{\Lambda}_{\psi s}(4459)$. The $\Xi_c\bar{D}(1/2^-)$ interaction generates an
S-wave bound state with a mass around 4335 MeV, closely matching the mass of
$P^{\Lambda}_{\psi s}(4338)$. The $\Xi'_c\bar{D}$ interaction produces a
molecular state with spin parity $1/2^-$. The $\Xi_c^*\bar{D}$ channel gives rise
to a molecular state with spin parity $3/2^-$. Even when the cutoff
parameter is adjusted to its maximum within a reasonable range, no bound states
are formed for the channels $\Lambda_c\bar{D}_s^{*}$ and $\Lambda_c\bar{D}_s$.

\subsection{Cross section for $J^P=1/2^-$ in energy region $[4350-4550]$~MeV}

In this section, we incorporate the couplings between the interaction channels
to study the kaon-induced production of $P^{\Lambda}_{\psi s}$. Since no
molecular state is found in the calculations without couplings near the
$\Lambda_c\bar{D}^*$ threshold, the energy regions can clearly be divided into
two parts. To present the results more effectively, we provide cross sections
for two distinct energy ranges: the higher energy region $[4350-4550]$~MeV and
the lower energy region $[4250-4380]$~MeV, where the LHCb observed states,
$P^{\Lambda}_{\psi s}(4459)$ and $P^{\Lambda}_{\psi s}(4338)$, are located,
respectively. Furthermore, to present results for the molecular pentaquarks with
different spin-parities, we focus on the partial wave cross sections rather than
the total cross sections.

\begin{figure}[h!]
  \centering
  \includegraphics[scale=0.61,bb=45 75 440 395,clip]{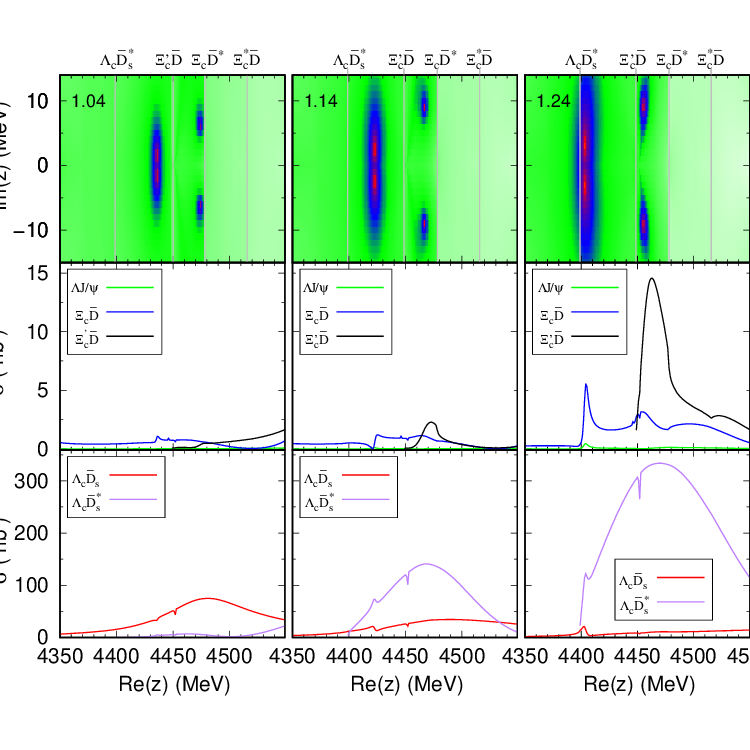}\\

\caption{The poles (upper panel) and the cross sections (middle and lower
panels) for spin parity $1/2^-$, calculated with different cutoffs of 1.04,
1.14, and 1.24~GeV. The gray lines in the upper panel correspond to the thresholds
of $\Xi^*_c\bar{D}$, $\Xi_c\bar{D}^*$, $\Xi'_c\bar{D}$, and
$\Lambda_c\bar{D}_s^{*}$, from right to left. In the middle and lower panels, the
curves represent the following channels: $\Lambda J/\psi$ (green),
$\Xi_c\bar{D}$ (blue), $\Xi'_c\bar{D}$ (black), $\Lambda_c\bar{D}_s$ (red), and
$\Lambda_c\bar{D}_s^*$ (purple).  }\label{Fig:4459.1}

\end{figure}

First, we consider the cross section for $J^P=1/2^-$ in the energy range
$[4350-4550]$~MeV. In the calculation without coupled-channel effects, two poles
with spin parity $1/2^-$ are generated in this energy region,
$\Xi'_c\bar{D}(1/2^-)$ and $\Xi_c\bar{D}^*(1/2^-)$, with the latter corresponding
to $P^{\Lambda}_{\psi s}(4459)$. The imaginary part of the pole with a cutoff of 1.14 GeV is about 10 MeV, corresponding to a width of approximately 20 MeV, which is close to the value reported by LHCb, 17 MeV~\cite{Aaij:2020gdg}. In Fig.~\ref{Fig:4459.1}, after including the
coupled-channel effects, the poles for these two states are shown in the complex
energy plane.  The corresponding cross sections in the channels considered are also
presented.

In the calculation without couplings, two poles are observed on the real axis,
positioned just below the thresholds of $\Xi'_c\bar{D}$ and $\Xi_c\bar{D}^*$,
respectively. When the coupled-channel effects are included, the poles move off
the real axis and form two conjugate poles in the complex energy plane. As the
parameter increases, the poles shift left along the real axis, indicating
stronger binding between the coupled channels. Compared to the results from our
previous calculation without the $\bar{K}N$ channel, the pole positions remain
similar, as expected, due to the large mass gap between the $\bar{K}N$ channel
and the energy region considered here.

The poles in the complex energy plane manifest as structures in the cross
section, which can be calculated from the scattering amplitudes along the real
axis. As shown in Fig.~\ref{Fig:4459.1}, the cross sections exhibit complex
structures due to the coupled-channel interactions in the considered energy
region. Generally, the cross sections for the $\Lambda_c\bar{D}_s$ and
$\Lambda_c\bar{D}_s^*$ channels are on the order of tens to hundreds of
nanobarns (nb), significantly larger than those for the $\Xi_c\bar{D}$ and
$\Xi'_c\bar{D}$ channels, which are on the order of several nb due to different
phase spaces and exchange particles as shown in
Table~\ref{flavor factor}. The cross section for the $\Lambda J/\psi$ channel is on the order of 1nb or less, which is consistent with predictions from other theoretical studies\cite{Clymton:2021thh,Paryev:2023icm,Paryev:2022zdx,Clymton:2022qlr}. However, it is significantly smaller than those of other channels, as the $\Lambda J/\psi$ final state cannot couple directly to the $\bar{K} N$ channel.

Although the cross sections for the $\Lambda_c\bar{D}_s$ and
$\Lambda_c\bar{D}_s^*$ channels are large, no obvious structures corresponding to
the poles in the complex energy plane (i.e., the molecular pentaquarks) are
observed. This suggests that the large cross sections are primarily due to
direct scattering rather than the formation of molecular states. As the cutoff
$\Lambda$ increases from 1.04 to 1.24~GeV, the cross section for the
$\Lambda_c\bar{D}_s$ channel decreases from about 100 nb to around 10 nb, while
the cross section for the $\Lambda_c\bar{D}_s^*$ channel increases from a few nb
to about 300 nb. Additionally, a dip is likely to occur around 4450 MeV,
attributed to the threshold effect of the $\Xi'_c\bar{D}$ channel.

For both the $\Xi_c\bar{D}$ and $\Xi'_c\bar{D}$ channels, the cross sections
become more prominent. An obvious peak can be seen in the $\Xi'_c\bar{D}$
channel, with a cross section around 15 nb, which corresponds clearly to the
pole below the $\Xi_c\bar{D}^*$ threshold related to the $P^{\Lambda}_{\psi s}(4459)$
observed at LHCb. However, as the cutoff decreases, the
peak diminishes rapidly, and the poles shift towards the $\Xi_c\bar{D}^*$
threshold. At a cutoff of 1.14 GeV, the peak cross section drops to just a few
nb, becoming nearly unobservable at a cutoff of 1.04 GeV. A relatively small
peak is also observed in the $\Xi_c\bar{D}$ channel at a cutoff of 1.24 GeV,
which transitions into a dip at a cutoff of 1.14 GeV, and then a very small peak
at a cutoff of 1.04 GeV. This structure originates from the pole below the
$\Xi'_c\bar{D}$ threshold, which also appears as a small peak in the
$\Lambda_c\bar{D}_s^*$ channel at cutoffs of 1.24 and 1.14 GeV.

\subsection{Cross section for $J^P=3/2^-$ in energy region $[4350-4550]$~MeV}

\begin{figure}[h!]
  \includegraphics[scale=0.61,bb=45 75 440 390,clip]{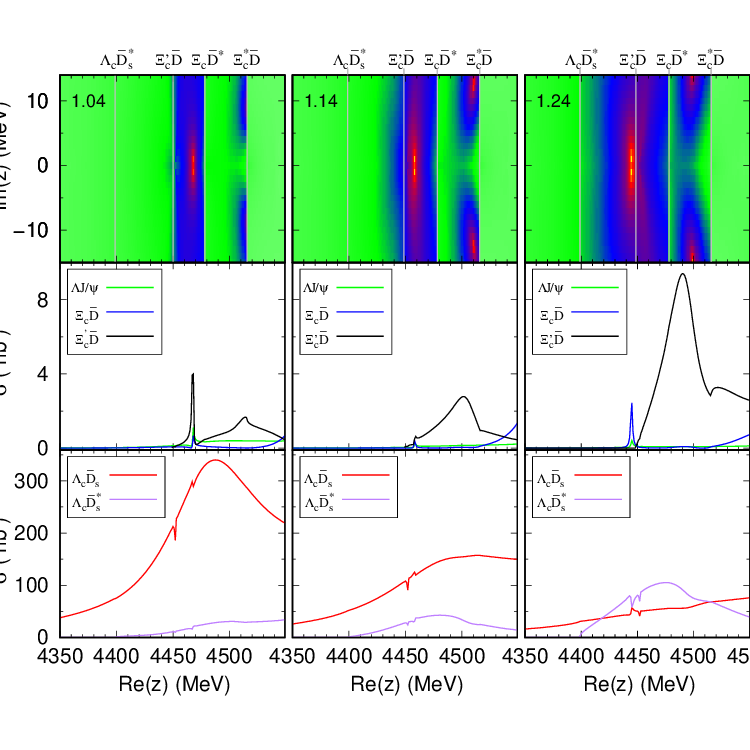}\\
  \caption{Similar to Fig.~\ref{Fig:4459.1}, but for the spin-parity $J^P = 3/2^-$.}\label{Fig:4459.2}
\end{figure}

Now, we consider another partial wave with spin parity $J^P = 3/2^-$ in the same
energy region. Without the couplings, two states with $J^P = 3/2^-$ are produced
near the $\Xi_c\bar{D}^*$ and $\Xi^*_c\bar{D}$ thresholds. After including the
coupled-channel effects, the two states move off the real axis and acquire a
width, as shown in Fig.~\ref{Fig:4459.2}. Similar to the $1/2^-$ case, the cross
sections for the channels $\Lambda_c\bar{D}_s$ and $\Lambda_c\bar{D}_s^*$ are on
the order of tens to hundreds of nb, which is much larger than the cross
sections for the $\Xi_c\bar{D}$ and $\Xi'_c\bar{D}$ channels, which are on the
order of a few nb. The cross section for the $\Lambda J/\psi$ channel is much
smaller than that for the other channels.

The higher state near the $\Xi^*_c\bar{D}$ threshold has a large width. With the
increase of the cutoff, the width becomes larger, and the pole in the complex
energy plane moves farther from the threshold. A peak corresponding to the pole
can be observed in the cross section of the $\Xi'_c\bar{D}$ channel, and with
the increase of the cutoff from 1.04 to 1.24 GeV, the peak becomes higher and
more pronounced. However, this pole does not appear in other channels.

The lower state near the $\Xi_c\bar{D}^*$ threshold, which can be related to the
$P^{\Lambda}_{\psi s}(4459)$ observed at LHCb, has a small width. At a cutoff of
1.04 GeV, the pole is close to the real axis and exhibits as a very narrow peak
in the cross section of the $\Xi'_c\bar{D}$ channel. However, with the increase
of the cutoff to 1.14 GeV, the pole moves to the $\Xi'_c\bar{D}$ threshold and
naturally can no longer be observed in the $\Xi'_c\bar{D}$ channel. When the
cutoff increases further to 1.24 GeV, the pole will even cross the
$\Xi'_c\bar{D}$ threshold, and the corresponding peak in the $\Xi_c\bar{D}$
channel becomes more obvious. This state also exhibits as a small dip in the
$\Lambda_c\bar{D}_s$ channel. Additionally, a dip is also found in the
$\Lambda_c\bar{D}_s$ channel near the $\Xi'_c\bar{D}$ threshold, as in the case
with $1/2^-$.

\subsection{Cross section for $J^P=1/2^-$ in energy region $[4250-4380]$~MeV}

The calculation without coupling effects suggests the presence of a bound state
originating from the $\Xi_c\bar{D}$ interaction, corresponding to
$P^{\Lambda}_{\psi s}(4338)$. As shown in Fig.4, after including coupled-channel effects, the state acquires a small imaginary part corresponding to a width of about 5 MeV, consistent with the LHCb observation~\cite{LHCb:2022ogu}. When the cutoff
parameter $\Lambda$ increases, the pole shifts leftward along the real axis, and
the pole near the $\Xi_c\bar{D}$ threshold becomes more prominent.

\begin{figure}[h!]
  \centering
  \includegraphics[scale=0.61,bb=45 165 440 390,clip]{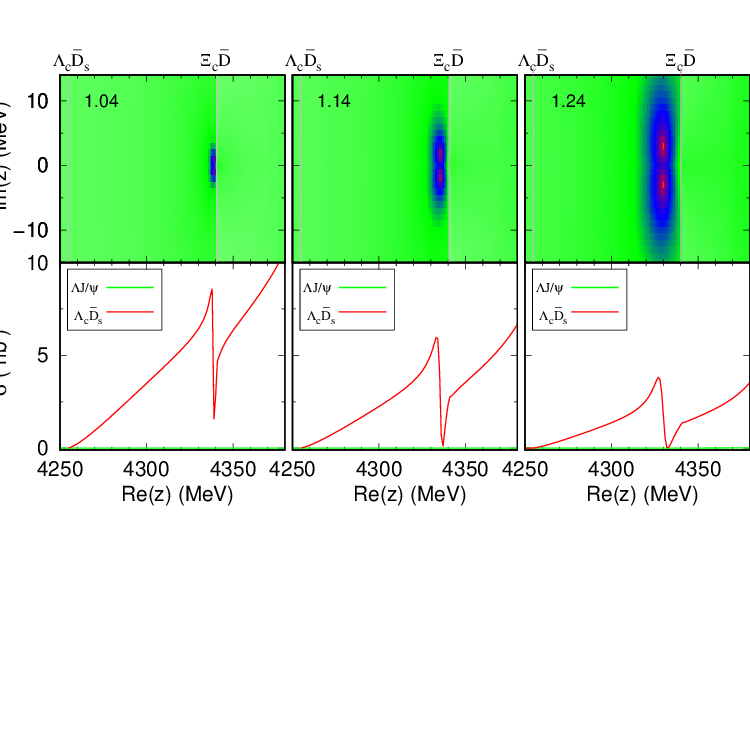}
  \caption{ Similar to Fig.~\ref{Fig:4459.1}, but in a lower energy region. }\label{mix16}
\end{figure}

Below $4338$ MeV, there are only two channels, the $\Lambda J/\psi$ and the
$\Lambda_c\bar{D}_s$, that we need to consider. From the calculation results of
the scattering cross section, regardless of the value of the parameter, the
scattering cross section in the $\Lambda J/\psi$ channel is very small and
almost undetectable. The scattering cross section results in the
$\Lambda_c\bar{D}_s$ channel show a swing structure occurring around $4340$ MeV at all
cuoffs of 1.04, 1.14 and 1.24~GeV. This is due to the pole near the $\Xi_c\bar{D}$ threshold, which corresponds to
$P^{\Lambda}_{\psi s}(4338)$, with interference from the background of other
interactions.

\section{Summary}\label{sum}

In this study, we investigated the kaon-induced production of strange
hidden-charm molecular pentaquark states, including $P^\Lambda_{\psi s}(4338)$
and $P^\Lambda_{\psi s}(4459)$, using a coupled-channel calculation. The
interactions considered involved the $\bar{K}N$, $\Xi_c^{(*)}\bar{D}^{(*)}$,
$\Xi'_c\bar{D}^{(*)}$, $\Lambda_c\bar{D}_s^{(*)}$, and $\Lambda J/\psi$ channels.
Effective Lagrangians were employed to construct the interaction potentials,
which were incorporated into the quasipotential Bethe-Salpeter equation to
compute scattering amplitudes. Partial wave cross sections for $J^P = 1/2^-$ and
$3/2^-$ are analyzed, revealing complex structures and significant
energy-dependent behavior.

For partial wave $J^P = 1/2^-$, in the lower energy region, the pole near the
$\Xi_c\bar{D}$ threshold, associated with $P^\Lambda_{\psi s}(4338)$, generates
a noticeable swing structure in the cross section for the $\Lambda_c\bar{D}_s$
and $\Lambda_c\bar{D}_s^*$ channels. This structure remains stable across
variations in the cutoff and exhibits a magnitude of several nb. In the
higher energy region, the $\Lambda_c\bar{D}_s$ and $\Lambda_c\bar{D}_s^*$
channels exhibit large cross sections, reaching up to several hundreds of
nb. However, no distinct peaks corresponding to molecular states were
observed, apart from threshold-induced cusps, suggesting that these
contributions arise primarily from direct scattering processes rather than
resonance effects.

In the $\Xi'_c\bar{D}$ channel, a prominent peak, associated with a pole below
the $\Xi_c\bar{D}^*$ threshold, is observed with a cross section of
approximately 15 nb. This peak is linked to the $P^\Lambda_{\psi s}(4459)$ state
observed by LHCb. As the cutoff decreases, the peak diminishes significantly,
becoming barely visible. In the
$\Xi_c\bar{D}$ channel, a smaller peak is evident at a cutoff of 1.24 GeV,
transitions into a dip at 1.14 GeV, and reappears as a very weak peak at 1.04
GeV. These structures originate from the pole below the $\Xi'_c\bar{D}$
threshold, which also gives rise to small peaks in the $\Lambda_c\bar{D}_s^*$
channel at higher cutoff values.

For the partial wave $J^P = 3/2^-$, the pole near the $\Xi_c\bar{D}^*$
threshold, associated with the $P^\Lambda_{\psi s}(4459)$, manifests as a peak
in the $\Xi'_c\bar{D}$ and $\Xi_c\bar{D}$ channels. At a cutoff of 1.04 GeV,
this pole exhibits a narrow width, remaining close to the real axis, and
produces a sharp peak in the $\Xi'_c\bar{D}$ channel. As the cutoff increases,
the pole shifts to the $\Xi'_c\bar{D}$ threshold and disappears from the
$\Xi'_c\bar{D}$ channel. At even higher cutoffs, the pole crosses the
$\Xi'_c\bar{D}$ threshold, resulting in a narrow peak in the $\Xi_c\bar{D}$
channel. This state also generates a small dip in the $\Lambda_c\bar{D}_s$
channel.  Additionally, states near the $\Xi_c^{*}\bar{D}$ threshold produce a broad
peak in the $\Xi'_c\bar{D}$ channel, characterized by a large width and
relatively weak sensitivity to the cutoff. Furthermore, a dip near the
$\Xi'_c\bar{D}$ threshold is observed in the $\Lambda_c\bar{D}_s$ channel,
similar to the behavior seen in the $J^P = 1/2^-$ case. These observations
highlight the intricate interplay between thresholds and poles, driven by
coupled-channel dynamics.

These findings suggest that the strange hidden-charmed pentaquarks, such as
$P^\Lambda_{\psi s}(4338)$ and $P^\Lambda_{\psi s}(4459)$, may potentially be
observed in kaon-induced production processes. There are still many alternative
assignments for the $P^\Lambda_{\psi
s}$~\cite{Wang:2021itn,Peng:2020hql,Feijoo:2022rxf}. Studying hidden-charmed
pentaquarks through new production mechanisms, from both theoretical and
experimental perspectives,  may also be beneficial for distinguishing these
different assignments.   High-precision experiments utilizing kaon beams at J-PARC and JLab would
be instrumental in confirming the existence of these states and exploring their
properties in detail.

\vskip 10pt

\noindent {\bf Acknowledgement} J. He is supported by the National Science
Foundation of China (Grant No. 12475080), and J. T. Zhu is supported by the
Start-up Funds of Changzhou University (Grant No. ZMF24020043) and the National
Science Foundation of China (Grant No. 12405090).

\noindent {\bf Data Availability Statement} This manuscript has no associated data or the data will not be deposited. [Authors' comment: This is a theoretical study and no external data are associated with this work.]

\end{document}